\documentclass{jpsj3}
\usepackage{graphicx}
\usepackage[usenames]{color}
\usepackage{amsmath}
\usepackage{amssymb}
\title{Pressure versus concentration tuning of the superconductivity\\ in Ba(Fe$_{1-x}$Co$_x$)$_2$As$_2$}
\author{Sandra Drotziger$^{1,2}$, Peter Schweiss$^{1}$, Kai Grube$^{1}$, Thomas Wolf$^{1}$, Peter Adelmann$^{1}$, Christoph Meingast$^{1}$, and Hilbert v. L\"{o}hneysen$^{1,2}$
}
\inst{$^1$ Karlsruhe Institute of Technology, Institut f\"{u}r Festk\"{o}rperphysik, PO Box 3640, D-76021 Karlsruhe, Germany \\ $^2$ Karlsruhe Institute of Technology, Physikalisches Institut, D-76128 Karlsruhe, Germany}
\abst{
In the iron arsenide compound BaFe$_2$As$_2$, superconductivity can be induced either by a variation of its chemical composition, e.g., by replacing Fe with Co, or by a reduction of the unit-cell volume through the application of hydrostatic pressure $p$. In contrast to chemical substitutions, pressure is expected to introduce no additional disorder into the lattice. We compare the two routes to superconductivity by measuring the $p$ dependence of the superconducting transition temperature $T_c$ of  Ba(Fe$_{1-x}$Co$_{x}$)$_2$As$_2$ single crystals with different Co content $x$. We find that $T_c(p)$ of underdoped and overdoped samples increases and decreases, respectively, tracking quantitatively the $T_c(x)$ dependence. To clarify to which extent the superconductivity relies on distinct structural features we analyze the crystal structure as a function of $x$ and compare the results with that of BaFe$_2$As$_2$ under pressure.
} 
\kword{iron pnictide superconductors, Co-doped BaFe2As2, superconductivity, high pressure}%
\begin{document}
\maketitle
\section{Introduction}
\label{intro}
In heavy-fermion compounds and cuprate perovskites unconventional superconductivity is observed close to magnetic order \cite{Monthoux_sc}. The heavy-fermion compounds are intermetallics composed of 4$f$ or 5$f$ elements with superconducting transition temperatures of typically less than a few Kelvin. The cuprate high-$T_c$ superconductors, on the other hand, are doped Mott-Hubbard insulators, composed of weakly coupled superconducting CuO$_2$ planes, and exhibit the highest $T_c$ so far known, with values of more than 100\,K.
Recently, a family of new superconductors based on iron-pnictide layers has been discovered that might bridge the gap between these two material classes. In particular the 122 iron arsenides, $A$Fe$_2$As$_2$ ($A$ = Ca, Sr, Ba), share the same tetragonal, ThCr$_2$Si$_2$-type crystal structure (space group $I4/mmm$) with the prototypical heavy-fermion superconductor CeCu$_2$Si$_2$ while their relatively high $T_c$ values, and the quasi-two-dimensional structure given by the weakly bonded, superconducting Fe$_2$As$_2$ layers are reminiscent of the cuprate perovskites. 
Superconductivity in 122 iron arsenides was first discovered in Ba$_{1-y}$K$_y$Fe$_2$As$_2$ \cite{Rotter1}. The parent compound BaFe$_2$As$_2$ exhibits collinear, antiferromagnetic spin-density-wave order below $T_{\rm{N}} \approx 140$\,K together with a structural transition to an orthorhombic crystal structure (space group $Fmmm$) \cite{Huang_122_mag}. When Ba is replaced with K these transitions split and are shifted to lower temperatures and superconductivity appears. With increasing K content, $T_c$ grows and reaches its maximum of 38\,K near the onset of magnetic order. In analogy to the heavy-fermion and cuprate superconductors, this has given rise to the conjecture that the superconducting pairing mechanism is essentially based on critical magnetic fluctuations. As a consequence, it was expected that any disorder should destroy the superconductivity, especially, if the superconducting gap has line or point nodes. An example for such a high sensitivity to impurities is the $d$-wave cuprate superconductor YBa$_2$Cu$_3$O$_{7-\delta}$ \cite{Jayaram_Dreck}. Here, already 5\% Zn in the superconducting CuO$_2$ planes suppress superconductivity completely. 
Unexpectedly, the substitution of Fe by Co in the 122 systems induces superconductivity with a qualitatively similar phase diagram as Ba$_{1-x}$K$_x$Fe$_2$As$_2$ but with a reduced $T_c$ maximum of 24\,K \cite{Sefat_Co_doped,Chu_PD}. 
As the substitution of K or Co introduces holes or electrons into the system, respectively, it was suggested that the charge carrier concentration controls the superconductivity, similar to the cuprate superconductors. High-pressure experiments on the antiferromagnetic parent compounds $A$Fe$_2$As$_2$ ($A$ = Ba, Sr) demonstrate, however, that---yet unidentified---structural changes alone are sufficient to induce superconductivity  \cite{Mani_pressure,Fukazawa_pressure,Alireza_pressure} resembling pressure induced superconductivity in heavy-fermion systems. 
Motivated by the fact that pressure does not introduce chemical disorder, in contrast to chemical substitutions, we investigated the pressure dependence of the super\-con\-ductivity in Ba(Fe$_{1-x}$Co$_x$)$_2$As$_2$ to dis\-entangle the effects of electron doping, structural changes, and disorder.
As distinguished from most of the published high-pressure investigations we used magnetization instead of transport measurements to be able to identify the thermodynamic signature of superconductivity. 
\begin{figure}[t] 
\includegraphics{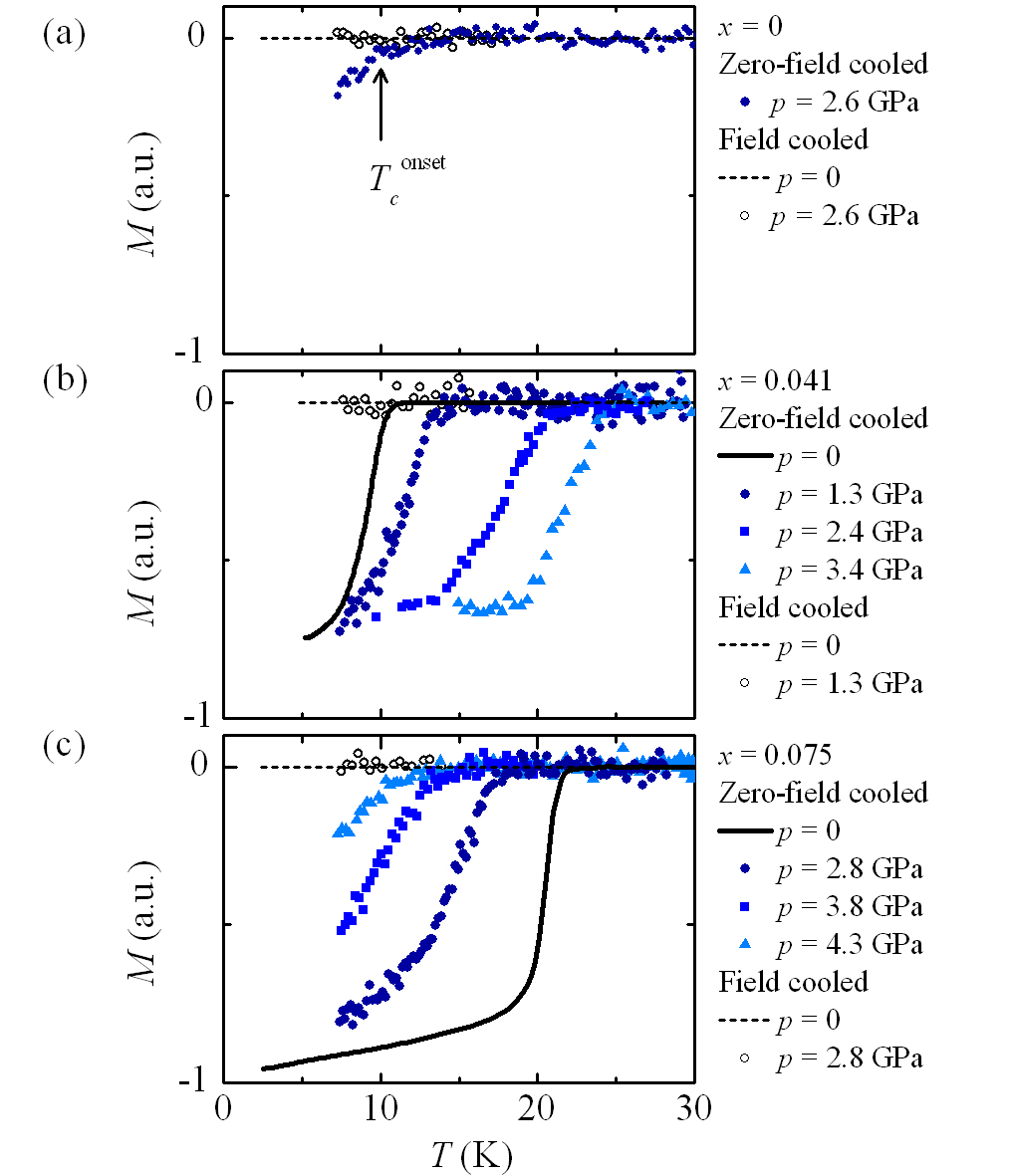}%
\caption{\label{fig:MT} (Color online) Zero-field-cooled and selected field-cooled magnetization measurements of Ba(Fe$_{1-x}$Co$_{x}$)$_2$As$_2$ singles crystals with $x=0$ (a), $0.041$ (b), and $0.075$ (c) in a magnetic field of $B=5\,$mT parallel to the $c$ axis.}
\end{figure}
\section{Sample Preparation and Experimental Methods}
\label{sec:exp}
Ba(Fe$_{1-x}$Co$_{x}$)$_2$As$_2$ single crystals were grown with a Fe-As self-flux method in alumina crucibles \cite{Hardy_expansion}. The actual Co concentration $x$ was determined by an XPS-microprobe analysis. To determine the crystal structure as a function of $x$ we used X-ray diffraction analysis with a four-circle diffractometer and Mo K$_{\alpha}$ radiation at room temperature and $p=0$. The subsequent structure refinement was performed with the aid of the SHELXS program. 
For the high-pressure magnetization measurements we built a miniaturized diamond-anvil cell that fits into a vibrating sample magnetometer (Oxford Instruments). The cell has an outer diameter of $12\,$mm and a length of $40\,$mm. To allow a maximum pressure of 10\,GPa the diamond anvils have a culet diameter of less than 0.8\,mm.
The pressure cell is made from an annealed CuBe alloy to minimize any magnetic contributions. Due to the extremely small sample signal, however, the background signal of the paramagnetic cell material and the toolmarks from the manufacturing process cannot be neglected and have to be determined by separate measurements of the empty cell. To provide quasi-hydrostatic pressure conditions we used Daphne Oil 7373 (Idemitsu Co., Japan) as pressure-transmitting medium. Due to the difference between the thermal expansion of the cell body and the anvils the applied pressure varies  by more than 10\% between room temperature and $4$\,K. Therefore, we used a Raman spectrometer with a $^4$He cooling stage to determine the pressure at $T_c(p)$ with the ruby-fluorescence method.
The superconducting properties of the Ba(Fe$_{1-x}$Co$_x$)$_2$As$_2$ crystals were first studied at ambient pressure in a Quantum Design SQUID magnetometer. The diamagnetic shielding and Meissner effect was investigated with zero-field-cooled (ZFC) and field-cooled (FC) magnetization measurements with a magnetic field of $5$\,mT parallel to the $c$ axis. Essentially, the Meissner signal (FC) was found to be negligible. From the crystals characterized in this way we cut small plate-like samples with typical dimensions of $50 \times 50 \times 20\,\mu$m$^3$ and inserted them into the diamond-anvil cell. The subsequent experiments were carried out in the same manner as the ambient pressure measurements. The masses of the samples differ typically by 20\%. 
Due to the difficulty to determine the exact mass of the samples used in the pressure cell it is impossible to give absolute magnetization values.
\begin{figure}[t]
\includegraphics{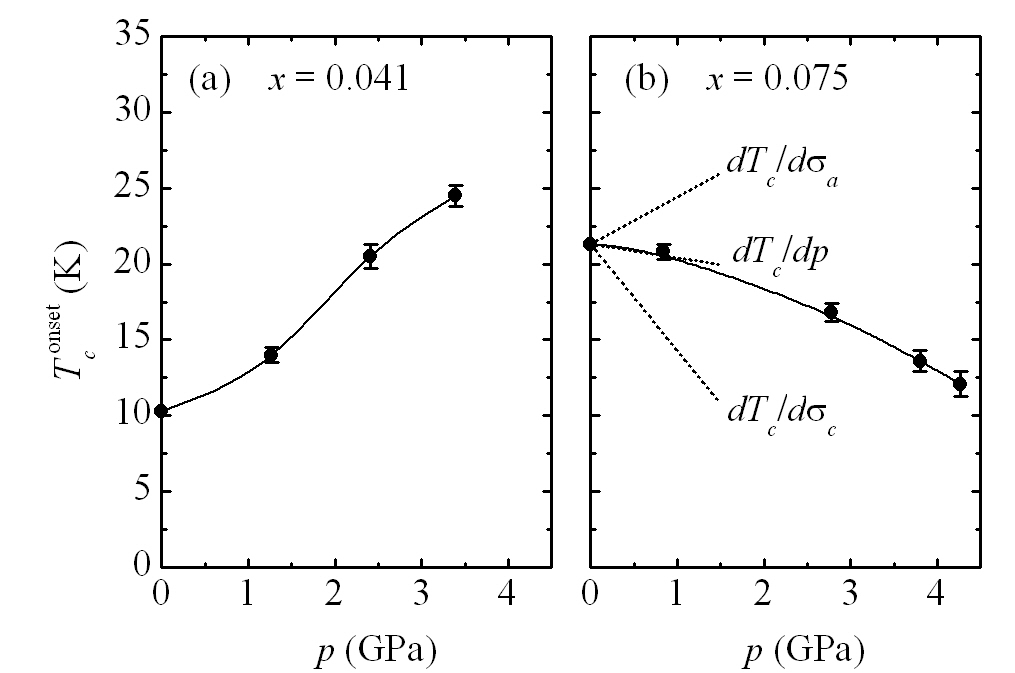}%
\caption{\label{fig:TcP} The magnetic onset of the superconducting transition temperature versus pressure of Ba(Fe$_{1-x}$Co$_{x}$)$_2$As$_2$ single crystals with $x=0.041$ (a) and $0.075$ (b). The continuous lines are guides to the eyes. The dotted lines show the initial uniaxial and hydrostatic pressure dependences $dT_c/d\sigma_{a}=3.1(1)\,$K/GPa, $dT_c/d\sigma_{c}=-7.0(2)\,$K/GPa, and $dT_c/dp=-0.9(3)$K/GPa, respectively, obtained from specific heat and thermal expansion measurements \cite{Hardy_expansion}.}
\end{figure}
\section{Magnetization Measurements}
\label{sec:mag}
The magnetization data of the Ba(Fe$_{1-x}$Co$_{x}$)$_2$As$_2$ single crystals with $x$~= 0, 0.041, and 0.075 are displayed in Fig.~\ref{fig:MT}. As mentioned before, in our measurements the Meissner effect is significantly smaller than the diamagnetic shielding. This behavior seems to be characteristic for many iron arsenides and is usually attributed to strong flux trapping, possibly enforced by the random Co distribution \cite{Chu_PD}. As mentioned above, the parent compound BaFe$_2$As$_2$ has a normal-conducting, antiferromagntic ground state. Figure~\ref{fig:MT}(a) illustrates that a pressure of $2.6(1)\,$GPa is sufficient to induce superconductivity with magnetically determined onset transition temperature of $T^\text{onset}_c\approx 10\,$K. The sample with Co content of $x=0.041$, shown in Fig.~\ref{fig:MT}(b), is already superconducting at $p$~=~0 with $T^\text{onset}_c \approx 11\,$K. Under pressure its transition broadens and shifts to higher temperatures while the discontinuity $\Delta M$ at $T_c$ remains roughly constant. The overdoped sample with $x=0.075$ and a $T^\text{onset}_c(p$~=~$0)$ of $\approx 21.5\,$K shows the opposite behavior: Here, $T_c$ drops with increasing pressure (see Fig.~\ref{fig:MT}(c)). 
The strong diamagnetic shielding of the $x$~= 0.041 and 0.075 crystals and the consistence of their $T^\text{onset}_c$ values with other thermodynamic measurements clearly indicate bulk superconductivity \cite{Hardy_expansion,Budko_C}, although---in particular at low Co concentrations---a  normal-conducting volume fraction cannot be ruled out. The broadening of the transitions at high pressures might be due to a small, slowly increasing uniaxial pressure contribution due to a gradual loss of the hydrostacity of the pressure-transmitting medium.

\begin{figure}[t]
\includegraphics{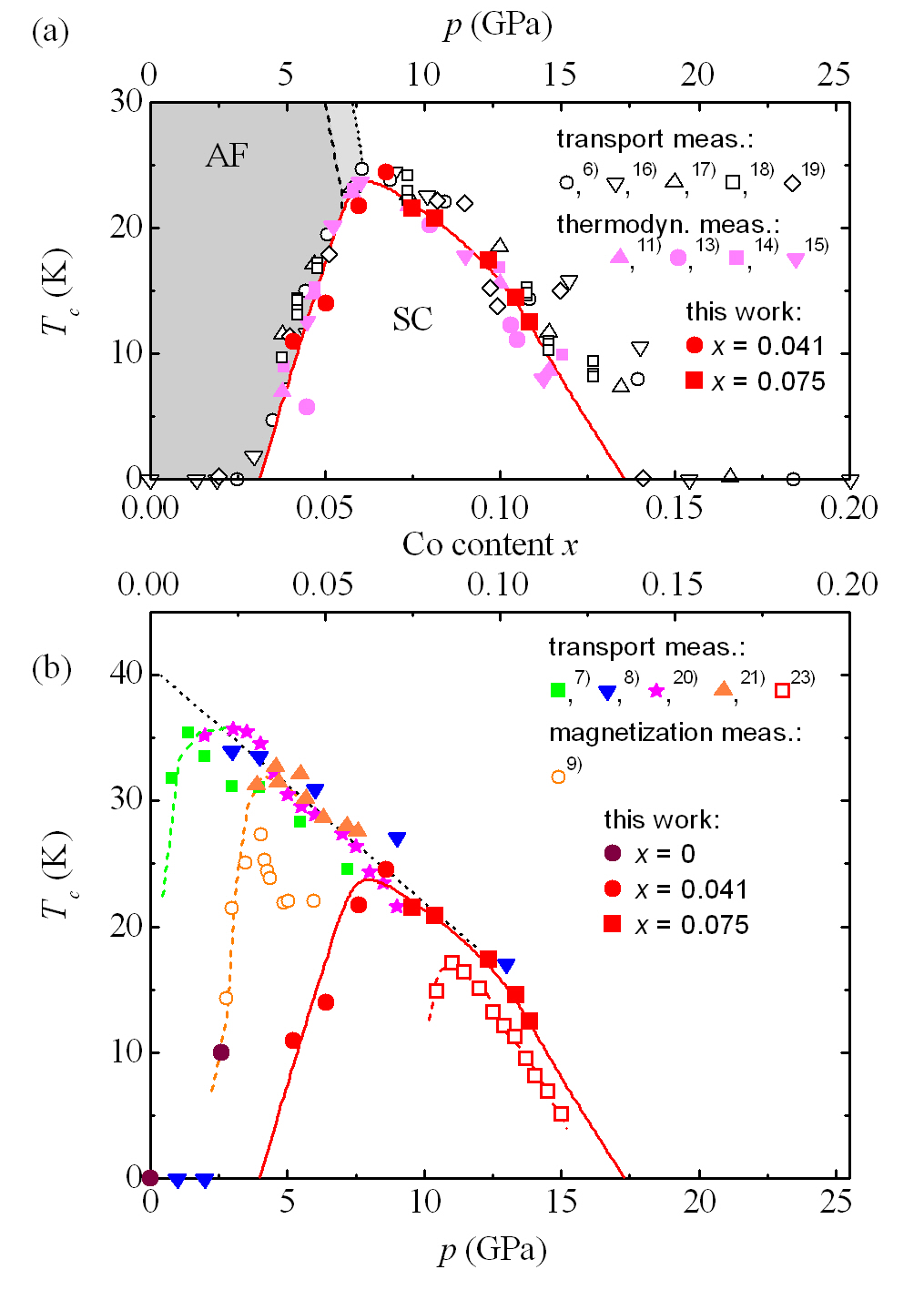}%
\caption{\label{fig:PD} (Color online) (a) Phase diagram of Ba(Fe$_{1-x}$Co$_{x}$)$_2$As$_2$ as a function of $x$ (lower scale) at $p$~=~0. The dotted and dashed lines denote the structural and magnetic transitions, respectively. At $x>0.06$, there is an increasing deviation between $T_c$ values determined by thermodynamic properties \cite{Budko_C,Gofryk_Cp,Prozorov_M_Co,Gang_cp_122} and transport measurements \cite{Chu_PD,Rullier_Hall,Canfield_doping,Reid_Rho_ab_Co,Ning_Rho_NMR_Co}. The $x$ values of the data from Ref.~\cite{Gang_cp_122} are scaled to match the $T_c$ maximum.
To compare the $T_c$ dependence on $x$ and $p$, we plot our $T_c(p)$ values of $x$ = 0.041 and 0.075 in the same phase diagram as a function of $p$ (upper scale) by assuming $\Delta p/\Delta x \approx 1.275\,$GPa/at.\%Co. The solid line is a guide to the eye. 
(b) Comparison of our measurements with high-pressure data of the undoped parent compound. These are resistivity \cite{Mani_pressure,Ishikawa_pressure,Fukazawa_pressure,Colombier_pressure,Yamazaki_pressure} and magnetization measurements \cite{Alireza_pressure}. The solid phase line $T_c(x,p=0)$ is taken from (a). The dashed lines illustrate the different, pressure-induced superconductivity onsets. The dotted line is a linear extrapolation of the high-pressure data to $p$~=~0.}
\end{figure}

In Fig.~\ref{fig:TcP} we summarize the measured $T^\text{onset}_c$ values as a function of $p$. The initial pressure dependence of the $x$~= 0.041 sample with $dT_c^\text{onset}/dp\approx 2.9(2)\,$K/GPa is supported by recent resistivity measurements which have been performed in a smaller pressure range on samples with different Co concentrations \cite{Ahilan_pressure_2}. At high doping levels, on the other hand, these measurements reveal nearly no change of $T_c$ under pressure in contrast to our measurement of the $x$~= 0.075 sample. With increasing Co content the $T_c$ values determined by transport and thermodynamic measurements start to deviate from each other, indicating minority superconducting phases with higher $T_c$ than the majority bulk phase of Co concentration $x$ (see the difference between the open and closed symbols in Fig.~\ref{fig:PD}(a)). These differences can be attributed to the sensitivity of resistivity measurements to filamentary superconductivity as opposed to bulk measurements of thermodynamic properties such as magnetization, thermal expansion, and specific heat.
Indeed, the initial slope $dT_c^\text{onset}/dp$ $=$ $-0.7(2)\,$K/GPa of the $x$ = 0.075 sample, obtained from our magnetization data, is convincingly confirmed by thermal expansion and specific heat measurements on a sample of the same batch. The Ehrenfest relations allow the determination of the uniaxial and hydrostatic pressure dependences of $T_c$ at $p$~=~0 from these data (see the dotted lines in Fig.~\ref{fig:TcP}(b)) \cite{Hardy_expansion}. In view of the strongly anisotropic uniaxial pressure dependences this excellent agreement proves that our data reflect indeed thermodynamic bulk properties, $dT_c^\text{onset}/dp=dT_c/dp$, under hydrostatic pressure conditions.

\begin{figure*}[t]
\includegraphics{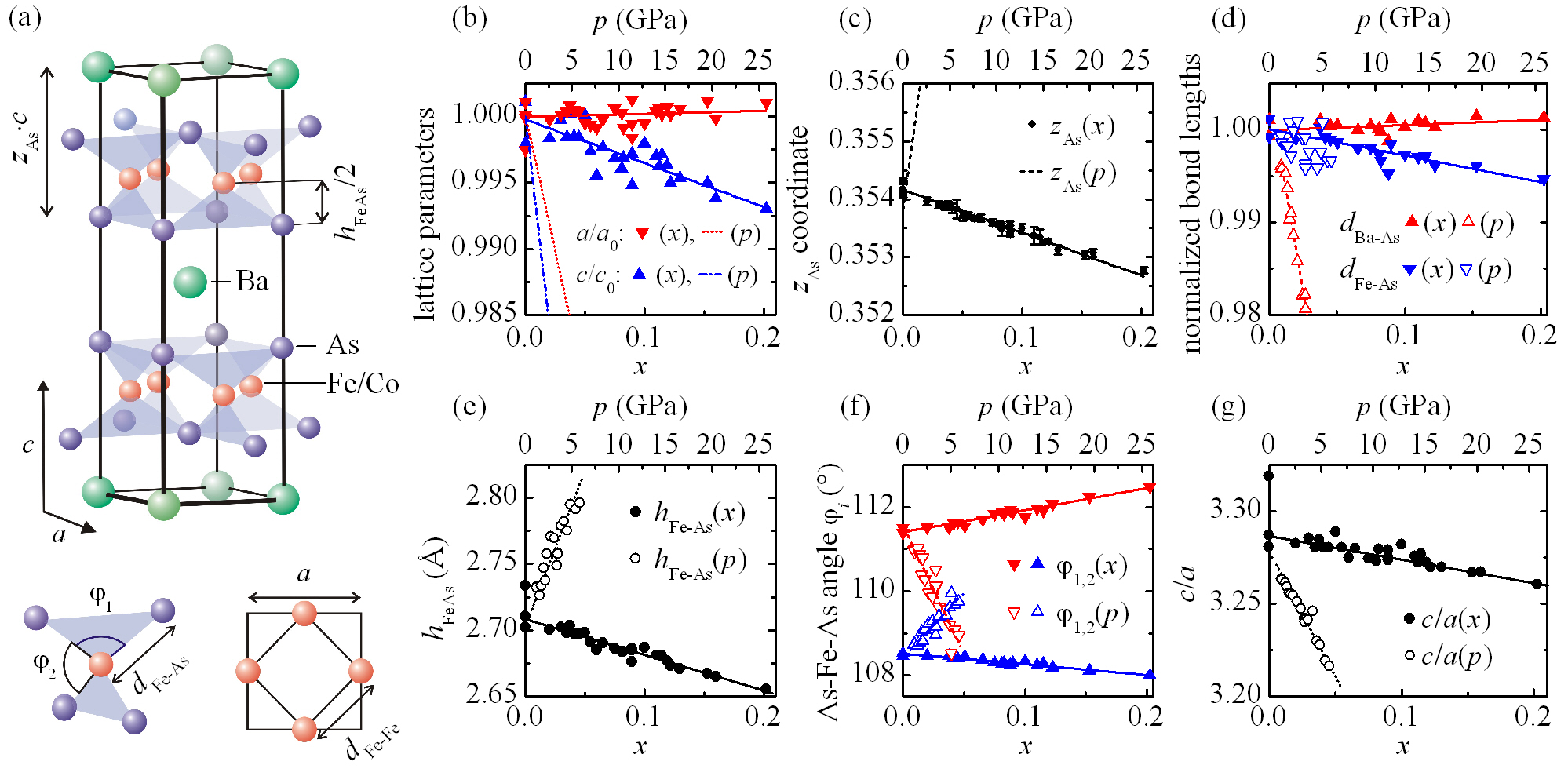}%
\caption{\label{fig:Struct} (Color online) (a) The tetragonal crystal structure of Ba(Fe$_{1-x}$Co$_x$)$_2$As$_2$. (b) The $a$ and $c$ axis divided by $a_0=3.966\,\text{\AA}$ and $c_0=13.037\,\text{\AA}$, respectively, and the As $z$ coordinate (c) as a function of $x$ (lower scale) at $p$~=~0. The broken lines represent the behavior of $a$, $c$, and $z_\text{As}$ of BaFe$_2$As$_2$ under pressure from Ref.~\cite{Kimber_pressure} plotted as a function of $p$ (upper scale) with the proportionality constant from Fig.~\ref{fig:PD}. (d) The Ba-As and (Fe$_{1-x}$Co$_x$)-As bond lengths divided by $3.374\,\text{\AA}$ and $2.392\,\text{\AA}$, respectively, the Fe$_2$As$_2$ layer thickness $h_\text{FeAs}$ (e), the As-Fe-As bond angles $\phi_i$ (f), and the $c/a$ ratio (d). The open symbols represent the corresponding quantities of BaFe$_2$As$_2$ under pressure \cite{Kimber_pressure}. All lines are linear fits to the data points.}
\end{figure*}

The $x$~= 0.041 concentration is at the underdoped and the $x$~= 0.075 at the overdoped side of the phase diagram where $T_c$ grows and drops with $x$, respectively. Hence, the sign change of $dT_c/dp$ mirrors that of $dT_c/dx$. To compare both effects quantitatively we assume that the $T_c$ change with $p$ is proportional to that with $x$ (see Fig.~\ref{fig:PD}(a)). Surprisingly, the data collapse on a single phase line if the proportionality constant is set to $\Delta p/\Delta x \approx 1.275\,$GPa/at.\%Co.
This scaling property of pressure and doping in the Fe$_2$As$_2$ planes is in remarkable contrast to the behavior found for cuprate superconductors where minute amounts of Zn in the CuO$_2$ planes quickly suppress superconductivity. On the other hand, a similar scaling of $T_c(p)$ and $T_c(x)$ at smaller doping levels was observed for Cd-doped CeCoIn$_5$ where Cd occupies the In sites \cite{Pham_Cd}. Substitution of magnetic and nonmagnetic ions into the Ce sublattice, however, leads likewise to a rapid reduction of $T_c$ \cite{Paglione_Ce115}.
A strong suppression of $T_c$ by nonmagnetic as well as magnetic impurities is a hallmark of unconventional, non-$s$-wave superconductivity \cite{Mackenzie_Dreck}. Usually, $T_c$ approaches zero when the charge carrier mean free path becomes smaller than the superconducting coherence length $\xi$. Recent $\mu$SR measurements demonstrate that at low Co content superconductivity develops in small islands around the randomly distributed Co ions \cite{Takeshita_insular}. For optimally doped Ba(Fe$_{1-x}$Co$_x$)$_2$As$_2$, $\xi$ is of the order of the $a$ axis lattice parameter and, hence, larger than the mean Fe-Fe nearest-neighbor distance $\approx a/\sqrt{2(1-x)}$ (see Fig.~\ref{fig:Struct}(a)) \cite{yamamoto_coherence_length,Yin_coherence_length}. The fact that samples with smaller Co concentration under pressure match those of larger $x$ at $p=0$, especially at the $T_c$ maximum, proves that chemical disorder does not affect the transition temperature which is consistent with fully gapped superconductivity.
It is instructive to compare the $T_c$ values of Co doped samples with those of the undoped parent compound BaFe$_2$As$_2$ under pressure (see Fig.~\ref{fig:PD}(b)). In comparison to the Co doped samples under pressure, the various published $T_c(p)$ data of pure BaFe$_2$As$_2$ differ strongly at low pressures. As pointed out by Duncan \textit{et al.} \cite{Duncan_uniaxial}, already tiny amounts of uniaxial pressure can suppress the magnetic order and shift the onset of superconductivity to lower pressures. Consequently, the degree of hydrostacity of the pressure-transmitting medium used has a crucial effect on the measurement. It is reassuring that our $x$~=~0 data coincides with those of Alireza \textit{et al.} \cite{Alireza_pressure} who used the same pressure medium. (Since pressure inhomogeneities tend to increase with applied pressure, the higher reproducibility of the experiments on Co-doped samples are partly a result of the lower maximal pressures needed to observe significant $T_c$ changes.) The extremely high sensitivity of the orthorhombic phase to stress, typical of heavily twinned crystals \cite{Grube_mono,Budko_pressure}, might be considerably reduced if the Co ions act as pinning centers for the twin boundaries \cite{Tanatar_twin,Chuang_Co_pin}. Indeed, all high-pressure experiments on BaFe$_2$As$_2$ that succeeded to suppress the twinned, orthorhombic phase show superconductivity with the same pressure dependence of $T_c$. In accordance with the observed scaling between $x$ and $p$, they all merge into the phase boundary $T_c(x\cdot \Delta p/\Delta x)$ of the doped samples at $p$~=~0. 
A similar decoupling of the magnetostructural and superconducting transitions together with a common phase boundary at high doping levels has been reported for Ba(Fe$_{1-x}$M$_x$)$_2$As$_2$ with M = Co, Ni, Rh, and Pd, if $T_c$ is plotted against the doped extra electron at the Fe/M site or the $c/a$ ratio \cite{Canfield_doping}.
Notably, in both examples the measurements tend to delineate a uniform phase line, irrespective of the different endpoints of magnetic order, as illustrated in Fig.~\ref{fig:PD}(b). Since the difference of endpoints does not result in a corresponding shift of the entire superconducting ``dome'', the superconductivity cannot originate from the magnetostructural instability but is rather expelled by the onset of antiferromagnetic order due to the strong competition of the different ground states.
If the $T_c$ dependence of the overdoped and optimally doped regime is extrapolated to $p$ (or $x$) = 0, the maximal $T_c$ would amount to 40\,K, a value which is close to the highest $T_c$ found in doped 122 iron arsenides \cite{Rotter1}.

\section{Search for Structural key parameters}
\label{sec:struct}
The fact that superconductivity can be induced by hydrostatic pressure without doping suggests that distinct structural parameters control the ground state, in loose analogy to the $f$-atom separation in some heavy-fermion superconductors. The currently most promising key parameters are the $c/a$ ratio \cite{Hardy_expansion,Canfield_doping}, the next nearest Fe-Fe distance $d_\text{Fe-Fe}=a/\sqrt{2}$ \cite{Zhao_angle,Miyazawa_a_axis}, the Fe$_2$As$_2$ layer thickness $h_\text{FeAs}$ (or pnictogen ``height'' $h_\text{FeAs}/2$) \cite{Singh_As_height,Mizuguchi_As_height,Kuroki_As_height,Kuchinskii_hFeAs}, and the As-Fe-As bond angles $\phi_i$ ($i$ = 1,2) of the tetragonal structure \cite{Zhao_angle,Lee_angle}. 
Theoretical studies pointed out that magnetically mediated superconductivity favors quasi-two-dimensional structures \cite{Monthoux_sc}. Indeed, most of the discovered unconventional superconductors are characterized by strong magnetic and electronic anisotropies and comprise layered, superconducting building blocks. The heavy-fermion superconductors based on the HoCoGa$_5$ structure show even a linear relationship between $T_c$ and the ratio of the tetragonal lattice parameters $c/a$ \cite{Bauer_dimensionality}. As already shown in Ref. \cite{Hardy_expansion}, the uniaxial pressure dependences of $T_c$, depicted in Fig.~\ref{fig:TcP}, support a similarly strong influence of the $c/a$ ratio on superconductivity in the 122 iron-arsenide superconductors. 
The As-Fe-As bond angles (and $h_\text{FeAs}$), on the other hand, are suggested to control the density of states at the Fermi level, with the highest $T_c$ observed for an ideal tetrahedral angle of $109.47\,^{\circ}$, where $\phi_1$ and $\phi_2$ become equal.
An example for the important effect of $\phi_i$ might be given by the hole-doped Ba$_{1-y}$K$_y$Fe$_2$As$_2$. This compound exhibits in addition a similar change of $\phi_i$ and $d_\text{Fe-Fe}$ under pressure and doping with K \cite{Kimber_pressure}. Together with its related compound Sr$_{1-y}$K$_y$Fe$_2$As$_2$, it shows an approximate correlation between $dT_c/dy$ and $dT_c/dp$ \cite{Torikachvili_pressure,Gooch_pressure}.

Based on the equivalence between $T_c(p)$ and $T_c(x \cdot \Delta p/\Delta x)$ found in Ba(Fe$_{1-x}$Co$_x$)$_2$As$_2$, we are now able to check the relevance of the suggested structural parameters for the electron-doped 122 compounds. For this we analyzed the crystal structure of Ba(Fe$_{1-x}$Co$_x$)$_2$As$_2$ as a function of $x$ at room temperature and $p$~=~0 and compare it with BaFe$_2$As$_2$ measured under pressure at $T$~=~150\,K \cite{Kimber_pressure}. The temperature difference between the data sets can be neglected because the thermal expansion is small compared to the pressure and doping dependent changes \cite{Kimber_pressure}.
The structure is fully characterized by the lattice parameters $a$, $c$, and the $z$ coordinate of the As ion. In Fig.~\ref{fig:Struct}(b) and (c) these parameters are plotted against $x$ and $p$ using the proportionality constant from above. 
In accordance to other Co-doped iron arsenides \cite{Marcinkova_Co}, both axes exhibit only small, gradual changes, demonstrating homogeneous solid Ba(Fe$_{1-x}$Co$_x$)$_2$As$_2$ solutions up to $x$~= 0.2. With increasing $x$, the $a$ axis remains nearly unchanged and the $c$ axis exhibits a slight shrinkage, which is exclusively caused by a decrease of $h_\text{FeAs}$, as indicated by the drop of $z_\text{As}(x)$ in Fig.~\ref{fig:Struct}(c). In contrast, pressure leads to a shortening of both axes and an increase of $z_\text{As}(p)$ \cite{Kimber_pressure,Jorgensen_pressure}. 
The dissimilar behavior of $z_\text{As}$ as a function of $p$ and $x$ originates from the different compressibilities of the Ba-As and (Fe$_{1-x}$Co$_x$)-As bonds. As shown by Fig.~\ref{fig:Struct}(d), Co substitution leads to a tiny decrease of the (Fe$_{1-x}$Co$_x$)-As bond $d_\text{Fe-As}$ while the Ba-As distance $d_\text{Ba-As}$ increases slightly. Under pressure, too, the Fe-As bond hardly changes but the weak Ba-As bond exhibits a pronounced reduction. The staggered structure of the incompressible Fe-As bonds forms ``Nuremberg scissors'' so that under hydrostatic pressure the compression of the Fe$_2$As$_2$ layer along the $a$ axis leads to an increase of the layer thickness parallel to $c$, as shown in Fig.~\ref{fig:Struct}(e). 
This has the additional effect that with growing $x$ the As-Fe-As angles reveal an increasing deviation from the ideal tetrahedral angle (see Fig.~\ref{fig:Struct}(f)). In contrast to the application of pressure or doping with K, Ba(Fe$_{1-x}$Co$_x$)$_2$As$_2$ exhibits its $T_c$ maximum for a structure that is far away from that of a regular Fe-As tetrahedron. 
Finally we turn to the $c/a$ ratio displayed in Fig.~\ref{fig:Struct}(g). Although both, $x$ and $p$ lead to a decrease of $c/a$, the slopes differ by nearly one order of magnitude. It has, however, taken into account that a change of $c/a$ might affect---apart from the effective dimensionality---the charge carrier density due to a simultaneous change of the bond angles $\phi_i$, as already suggested in Ref.~\cite{Hardy_expansion}. To disentangle both effects we compare the $T_c$ values of different Co-doped 122 compounds $A$(Fe$_{1-x}$Co$_x$)$_2$As$_2$ with $A$~= Ca \cite{Kumar_Ca122}, Sr \cite{Leither_Sr122}, Eu \cite{Ying_Eu122_doped}, and Ba. The different $A$ ion radii result in a variation of their $c/a$ ratio that ranges from 3.01 to 3.28 for $A$~= Ca and Ba, respectively. All mentioned Co-doped compounds show the $T_c$ maximum approximately at the same Co concentration and, consequently, at the same charge carrier concentration. Therefore, the $T_c$ maximum as a function of $c/a$ should reflect the dependence on the effective dimensionality. Compared to the application of pressure and doping with Co, which show a $T_c$ dependence of $dT_c(p)/d(c/a)\approx 200\,$K and $dT_c(x)/d(c/a)\approx 1400\,$K, respectively, the in this way determined variation of $T_c$ at a fixed charge-carrier concentration is insignificantly small $dT_c(A)/d(c/a) \approx 13\,$K. This observation corresponds to the aforementioned electron-doped
Ba(Fe$_{1-x}M_x$)$_2$As$_2$ ($M$ = Co, Rh, Ni, Pt). It demonstrates that $c/a$ is predominately determined by the additional electrons doped at the Fe/$M$ site. Obviously, in contrast to our expectation, the interlayer distance and hence the dimensionality have nearly no effect on superconductivity. 

As a result neither the Fe-Fe distance $d_\text{Fe-Fe}$, nor the bond angles $\phi_i$, nor the Fe$_2$As$_2$ layer thickness (pnictogen height) $h_\text{Fe-As}$, nor the $c/a$ ratio meet the criteria for structural key parameters. 
The only parameter which might show a similar behavior with $x$ and $p$ is the Fe/Co-As bond length $d_\text{Fe-As}$, although additional experimental studies are necessary to prove whether the $d_\text{Fe-As}$ exhibits a comparable slight reduction under high pressure as with large Co concentrations. Recently, the importance of $d_\text{Fe-As}$ was pointed out by crystal structure investigations of electron and hole-doped 122 compounds which indicate that superconductivity favors a distinct Fe/$M$-As bond length \cite{Kim_FeAs}. First principle calculations show that $d_\text{Fe-As}$ determines the local magnetic moment on the Fe site \cite{Johannes_FeAs}. 
Due to its magnetostrictive nature a vanishing moment would be reflected in a pronounced reduction of $d_\text{Fe-As}$. In contrast to the experimentally determined values, the first-principle calculations predict for the optimized structure, without accounting for magnetism, a clearly smaller Fe/Co-As bond length. The fact that $d_\text{Fe-As}$ is large was taken as a hint for large magnetic moments and frustrated magnetic interactions \cite{Yildirim_Fe-As_bond,Mazin_Fe-As_nature}. As in addition $d_\text{Fe-As}$ exhibits only minor changes with increasing $x$ and $p$, even if the system reveals no longer magnetic order, the superconductivity has to evolve from a paramagnetic phase with strong magnetic fluctuations. In all iron-arsenides discovered so far the temperature of the magnetic transition is equal or smaller than that of the structural transformation. Apparently, the orthorhombic distortion is a prerequisite for long-range magnetic order, possibly due to the frustration of two antiferromagnetic sublattices \cite{Yildrim_AFM_ortho}. Taking this into account the phase diagrams depicted in Figure~\ref{fig:PD}(b) show that as soon as the structural transition is suppressed by pressure superconductivity replaces antiferromagnetism. Therefore, both, magnetic order and superconductivity seem to originate from the same, presumably magnetic interactions.

\section{Summary}
In conclusion we found a scaling of the phase diagram of Ba(Fe$_{1-x}$Co$_x$)$_2$As$_2$ with electron doping by Co or pressure. This gives rise to the assumption that distinct structural parameter are essential in achieving superconductivity. A detailed comparison of the key elements suggested so far reveals, however, a different, often even opposite, evolution of these parameters under pressure and Co-doping.
The only exception might be given by the Fe-As bond length. Its insensitivity to $p$ and $x$, however, requires additional high-resolution crystal-structure investigations to demonstrate a clear correlation to superconductivity. The discovered similarities should be useful to discriminate between different theoretical models. In this context, the study of 122 iron arsenides with other chemical substitutions under pressure and additional studies of the anisotropic uniaxial pressure dependence of $T_c$ would be helpful. 
The insensitivity of superconductivity to any chemical disorder clearly points to a nodeless superconducting gap. 
The decoupling of the magnetostructural and superconducting transitions and the uniform phase boundary at high pressure or high doping levels indicate a strong competition of the ground states and disfavor the antiferromagnetic quantum phase transition as source for superconductivity. The rigid Fe-As bond, on the other hand, makes clear that even outside the antiferromagnetic phase magnetic interactions are present and that magnetic order and superconductivity might have a common origin in the 122 iron arsenides.

\section*{Acknowledgments}
We acknowledge financial support through the Helmholtz association (VIRQ VH-VI-127) and the Deutsche Forschungsgemeinschaft (SPP 1458).
%
%
%

%
%

\end{document}